\documentclass[useAMS,usenatbib,letters]{mn2e}
\usepackage{journal_abbreviations}
\usepackage{graphicx, amsmath, amssymb}
\usepackage{enumerate}
\usepackage{float}

\usepackage{mathrsfs}
\usepackage{graphics}

\title[Cumulative number density in EAGLE]{A large difference in the progenitor masses of active and passive galaxies in the EAGLE simulation}
\author[Clauwens et al.]{Bart Clauwens$^{1,2}$\thanks{E-mail: clauwens@strw.leidenuniv.nl}, Marijn Franx$^{1}$, Joop Schaye$^{1}$\\
$^{1}$Leiden Observatory, Leiden University, PO Box 9513, 2300 RA Leiden, The Netherlands\\$^{2}$Instituut-Lorentz for Theoretical Physics, Leiden University, 2333 CA Leiden, The Netherlands\\}

\begin{document}

\date{Accepted 2016 July 11. Received 2016 June 23; in original form 2016 April 29.}

\pagerange{\pageref{firstpage}--\pageref{lastpage}}

\maketitle

\label{firstpage}

\begin{abstract}

Cumulative number density matching of galaxies is a method to observationally connect descendent galaxies to their typical main progenitors at higher redshifts and thereby to assess the evolution of galaxy properties. The accuracy of this method is limited due to galaxy merging and scatter in the stellar mass growth history of individual galaxies. \citet{Behroozi13} have introduced a refinement of the method, based on abundance matching of observed galaxies to the Bolshoi dark-matter-only simulation. The EAGLE cosmological hydro-simulation is well suited to test this method, because it reproduces the observed evolution of the galaxy stellar mass function and the passive fraction. We find agreement with the \citet{Behroozi13} method for the complete sample of main progenitors of $z=0$ galaxies, but we also find a strong dependence on the current star formation rate. Passive galaxies with a stellar mass up to $\rm{10^{10.75}M_{\odot}}$ have a completely different median mass history than active galaxies of the same mass. This difference persists if we only select central galaxies. This means that the cumulative number density method should be applied separately to active and passive galaxies. Even then, the typical main progenitor of a $z=0$ galaxy already spans two orders of magnitude in stellar mass at $z=2$.
\end{abstract}


\section{Introduction}
\label{SectionIntroduction}

Observations provide properties of samples of galaxies at different redshifts. Inferring the typical evolution of  individual galaxies from these observations is non-trivial, since it involves linking representative progenitor- and descendant-galaxies. This link is not directly observable, since every galaxy is normally only observed at one instance in time.

Once we have the ideal cosmological simulation that reproduces the evolution of all properties of galaxy samples across cosmic time, we can retrieve typical galaxy evolution tracks from this. However, at the moment, the space of possible simulations is many-dimensional and not well constrained by physics from first principles. Changes in one of the modelling assumptions can affect many predictions in a complicated way. Vice versa, the comparison of simulation predictions and observations does not easily translate into a required change in the model ingredients. For that reason, apart from comparing galaxy samples in observations and simulations, it is important to attempt to infer the typical evolution of individual galaxy properties as much as possible directly from observations.

Cumulative number density matching of galaxies across redshift is a promising method to achieve this. In its original form it does not need any simulation input. The cumulative number density at a given redshift and mass is defined as the comoving number density of galaxies with a stellar mass larger than or equal to the given mass. Main progenitors are then selected at a constant cumulative number density. The underlying assumption is that galaxies evolve conjointly, building up stellar mass in a similar way, without changing rank order (based on stellar mass or velocity dispersion).

This method, originating from the work of \citet{Loeb03}, has been employed by \cite{Papovich11} and \citet{Lundgren14} to study stellar mass and star formation rate (SFR) evolution out to $z\approx8$, by \citet{Dokkum10} and \citet{Patel13} to study the evolution of the structural parameters of massive galaxies out to $z\approx 3$, by \citet{Dokkum13} and \citet{Morishita15} to study the stellar density profile evolution of Milky-Way-like and massive galaxies since $z\approx3$ and by \citet{Finkelstein15} to predict the abundance of bright $z\approx9$ galaxies.

Cumulative number density matching is not expected to be a perfect method for inferring the evolution of galaxies, because it neglects galaxy mergers and because rank order may not be conserved. The viability of the method has therefore been investigated by \citet{Leja13}, applied forward in time to the descendants of $z\approx3$ galaxies, based on the  \citet{Guo11} semi-analytic model of galaxy formation. They find that a constant cumulative number density is a good first order approximation for these descendants. \citet{Behroozi13} apply the method backwards in time to the main progenitors of $z=0$ galaxies. They use by construction a representative history of the galaxy stellar mass function (GSMF), based on the abundance matching of observed galaxies to the Bolshoi dark-matter-only simulation. They find that a constant cumulative density is a poor prescription for matching main progenitors and they give a recipe to account for the increase in the running median cumulative number density towards higher redshifts that results from merging:  $(0.16 \Delta z)$ dex. This equation applies to a large range of galaxy masses and redshifts up to 8. Other recent studies have been undertaken by \citet{Torrey15b} based on the Illustris hydrodynamic simulation as well as by \citet{Mundy15},  \citet{Henriques15} and \citet{Terrazas16} for different semi-analytic methods.

In this work we investigate the accuracy of the cumulative number density matching technique by comparing to results of the EAGLE hydrodynamic simulation \citep{Schaye15,Crain15}. EAGLE is arguably the first hydrodynamic simulation that has an accurate enough evolution of the GSMF \citep{Furlong15} and a representative enough passive/active galaxy population \citep{Schaye15,Trayford16} to address this question in some detail.

\section{Simulation}
\label{SectionSimulation}

We follow the main progenitors of redshift zero galaxies in the $\rm{(100 \; Mpc)^3}$ sized EAGLE simulation RefL100N1504. This simulation has been calibrated to the $z=0$ GSMF and mass-size relation. It has an initial gas particle mass of $\rm{1.8\times10^6\; M_{\odot}}$ and a maximum gravitational force softening of 700 pc. We use the public data release described in \citet{McAlpine16}. Following \citet{DeLucia07}, the main progenitor is defined as the progenitor with the most massive integrated history.

\section{Results}
\label{SectionResults}

\begin{figure*}
\includegraphics[height=0.2794\textwidth]{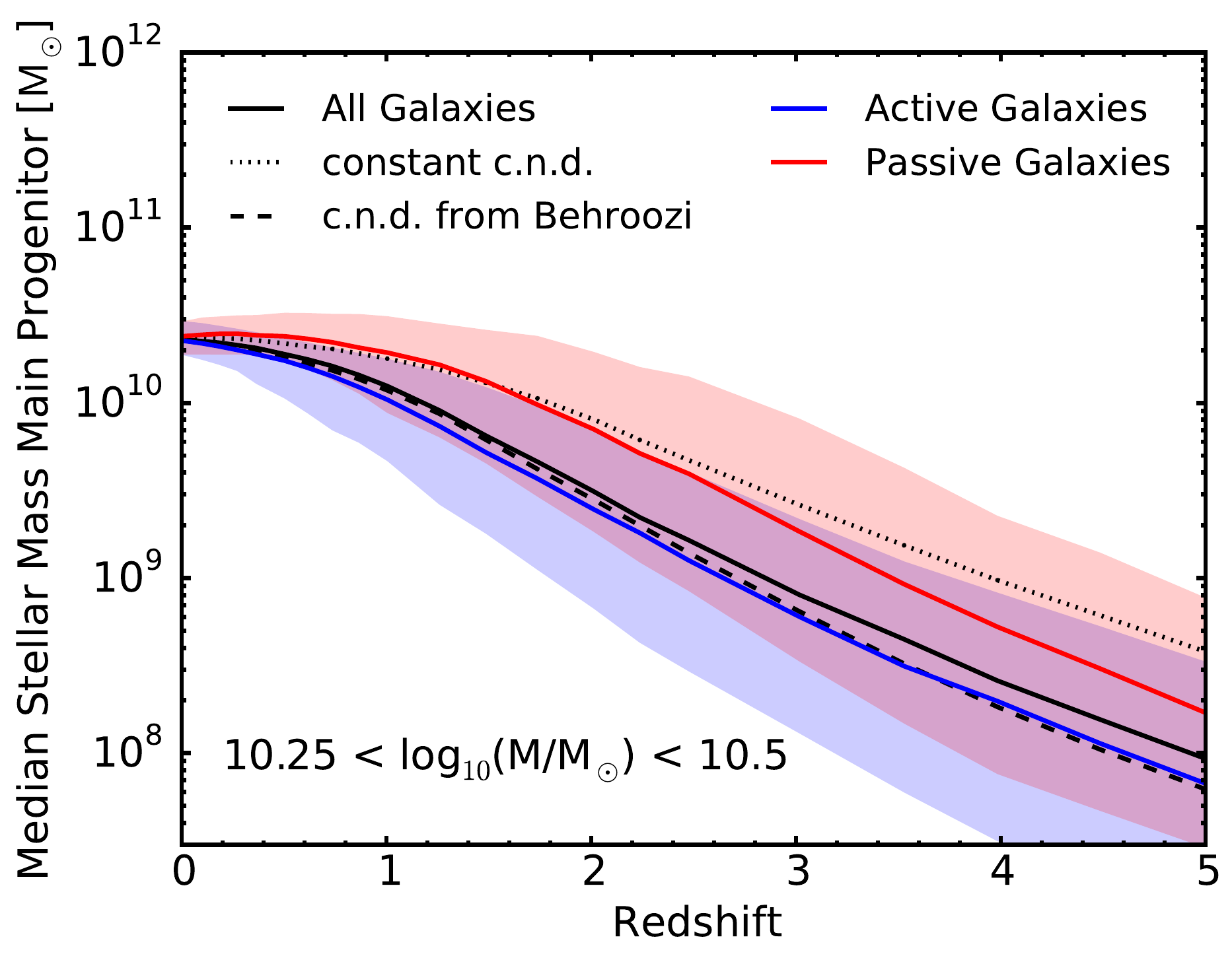}
\includegraphics[height=0.27\textwidth]{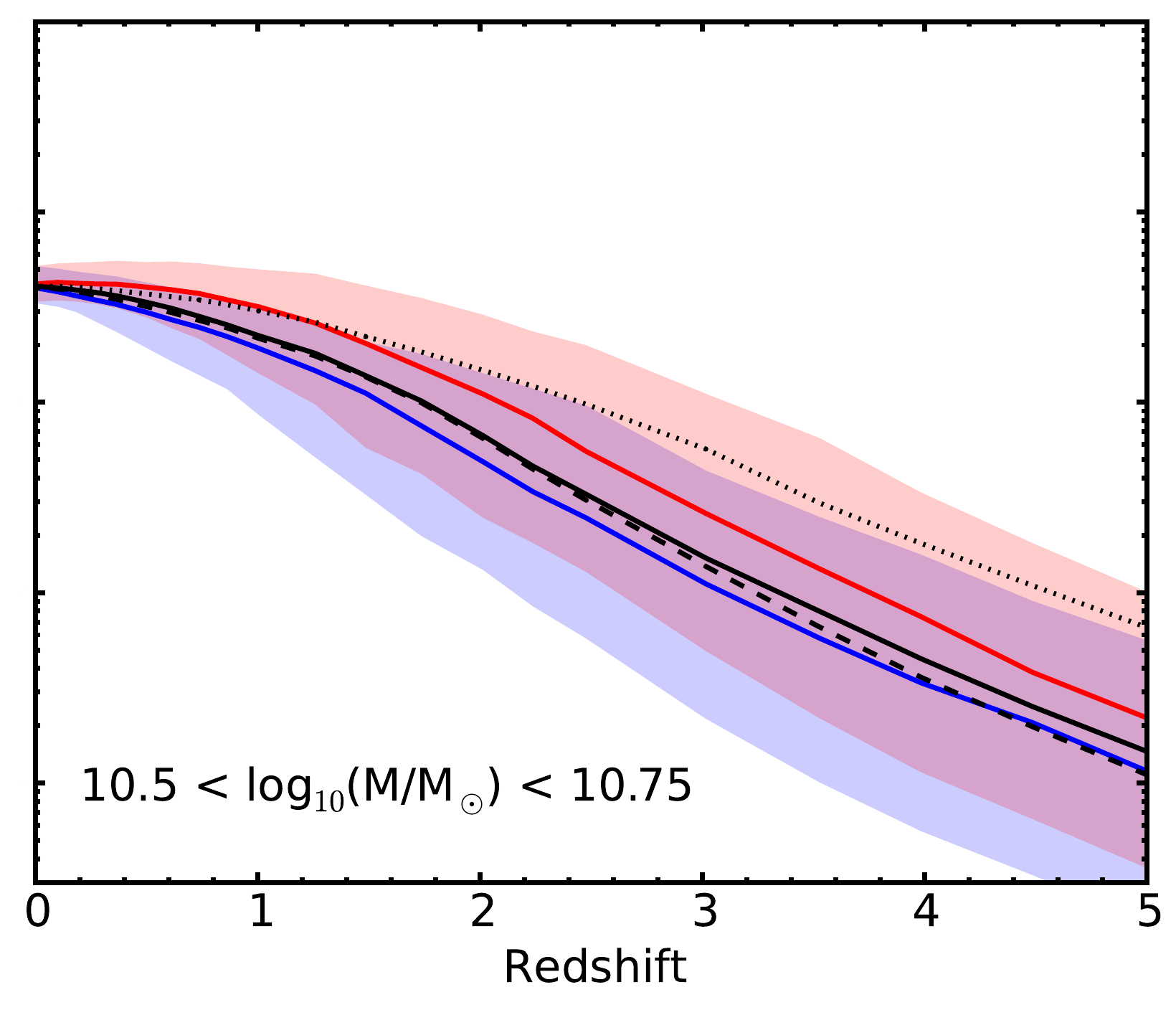}
\includegraphics[height=0.27\textwidth]{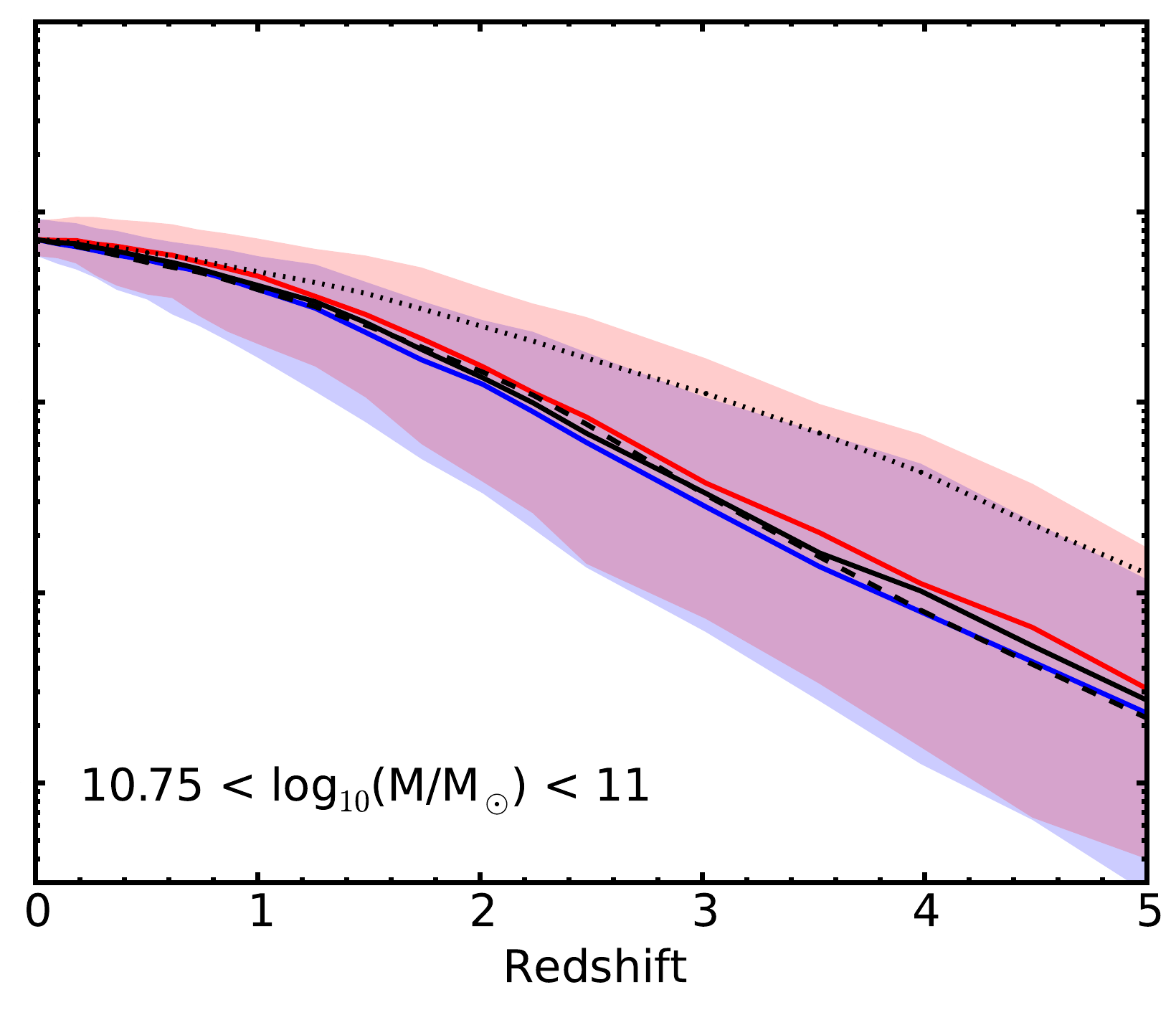}
\caption{The history of the median main progenitor stellar mass in the EAGLE simulation for three 0.25 dex wide mass bins, selected at $z=0$. The solid black curve denotes the true median main progenitor mass. The dotted black curve shows the stellar mass at a constant cumulative number density. The dashed black curve follows the stellar mass at an exponentially evolving cumulative number density as suggested by \protect{\citet{Behroozi13}}. Solid blue and red curves denote the true median main progenitor masses for the subsets of $z=0$ active respectively passive galaxies. The shaded regions denote the corresponding $\rm{10^{th}-90^{th}}$ percentiles.}
\label{Figure1}
\end{figure*}

\begin{figure*}
\includegraphics[height=0.2757\textwidth]{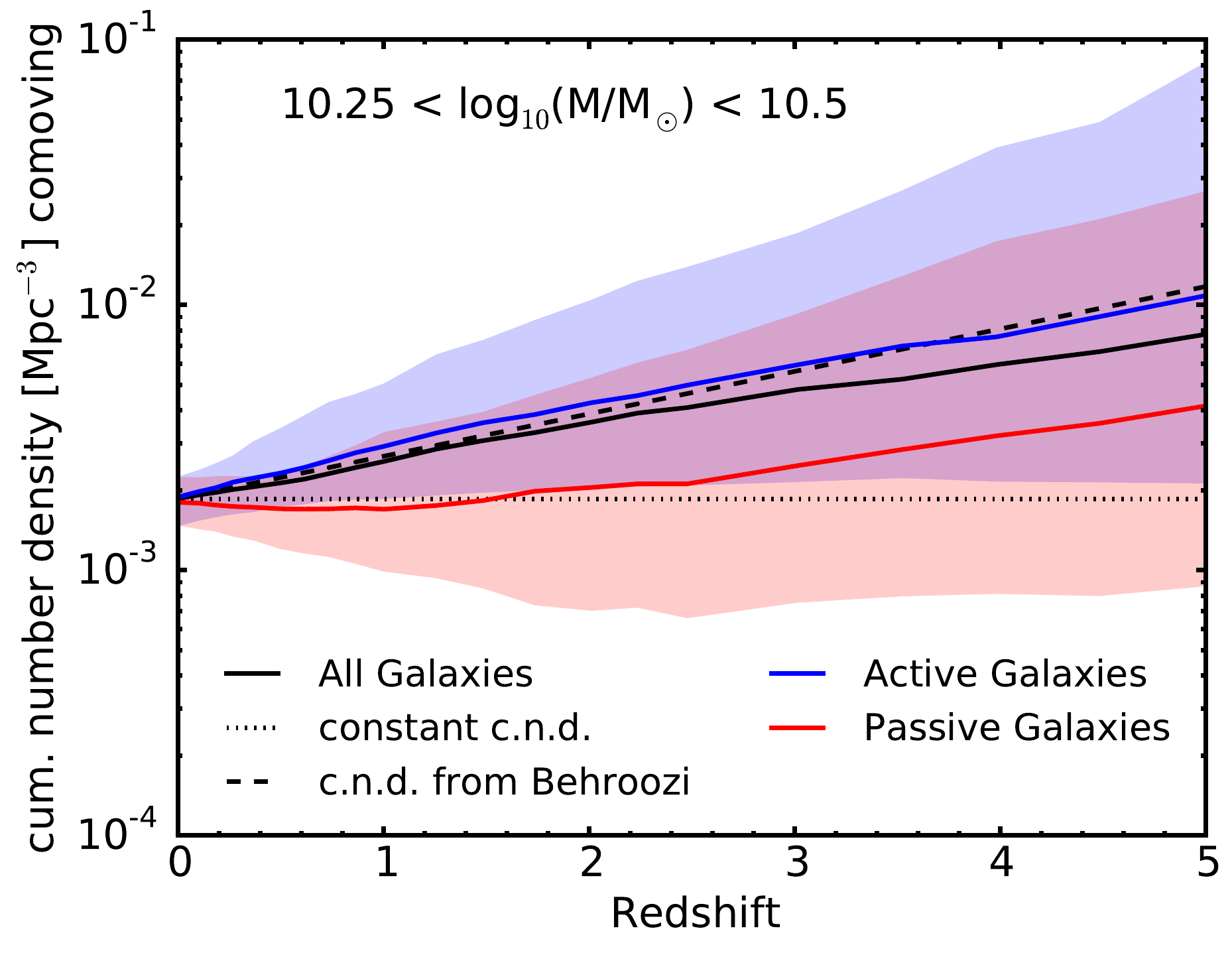}
\includegraphics[height=0.27\textwidth]{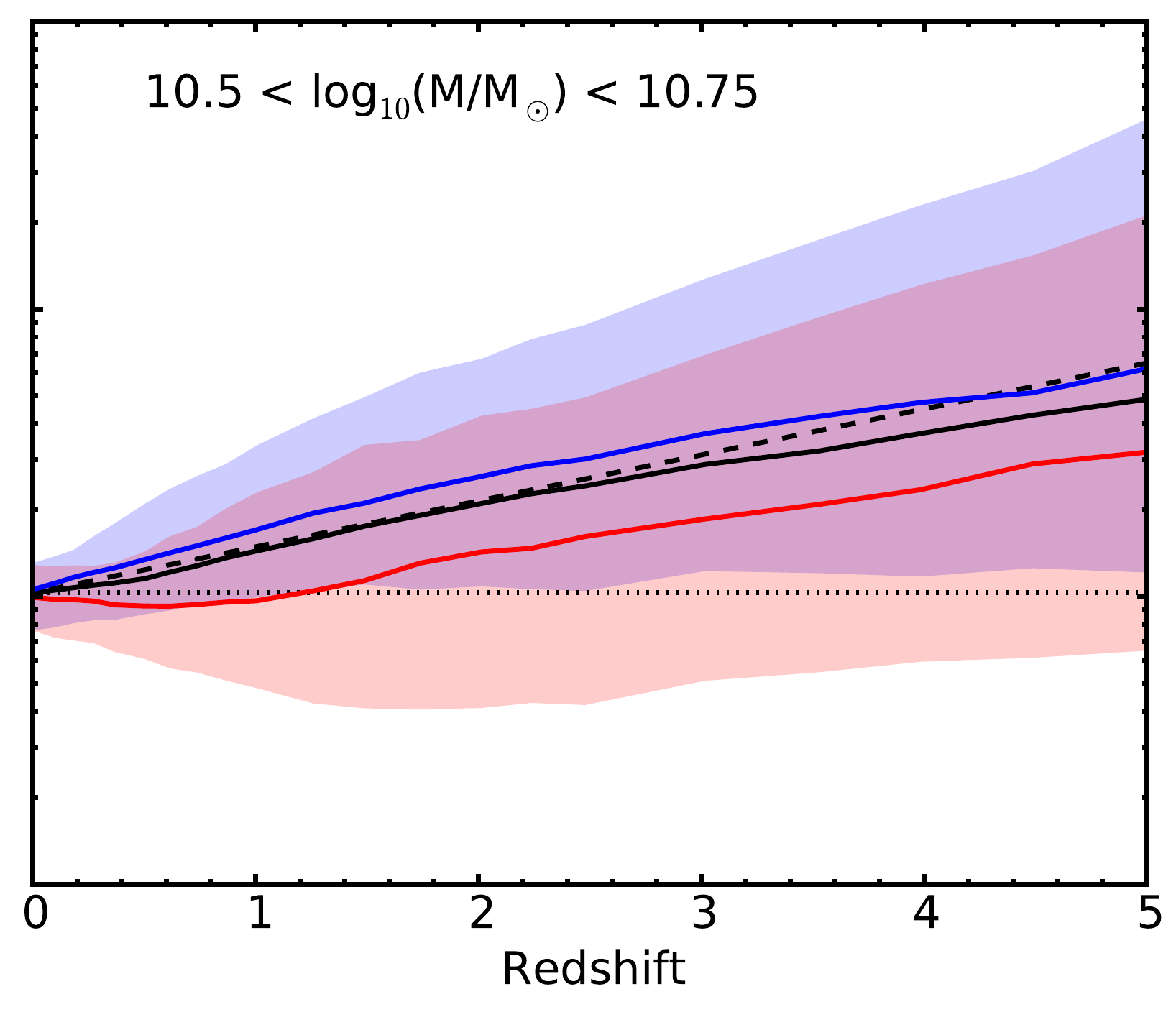}
\includegraphics[height=0.27\textwidth]{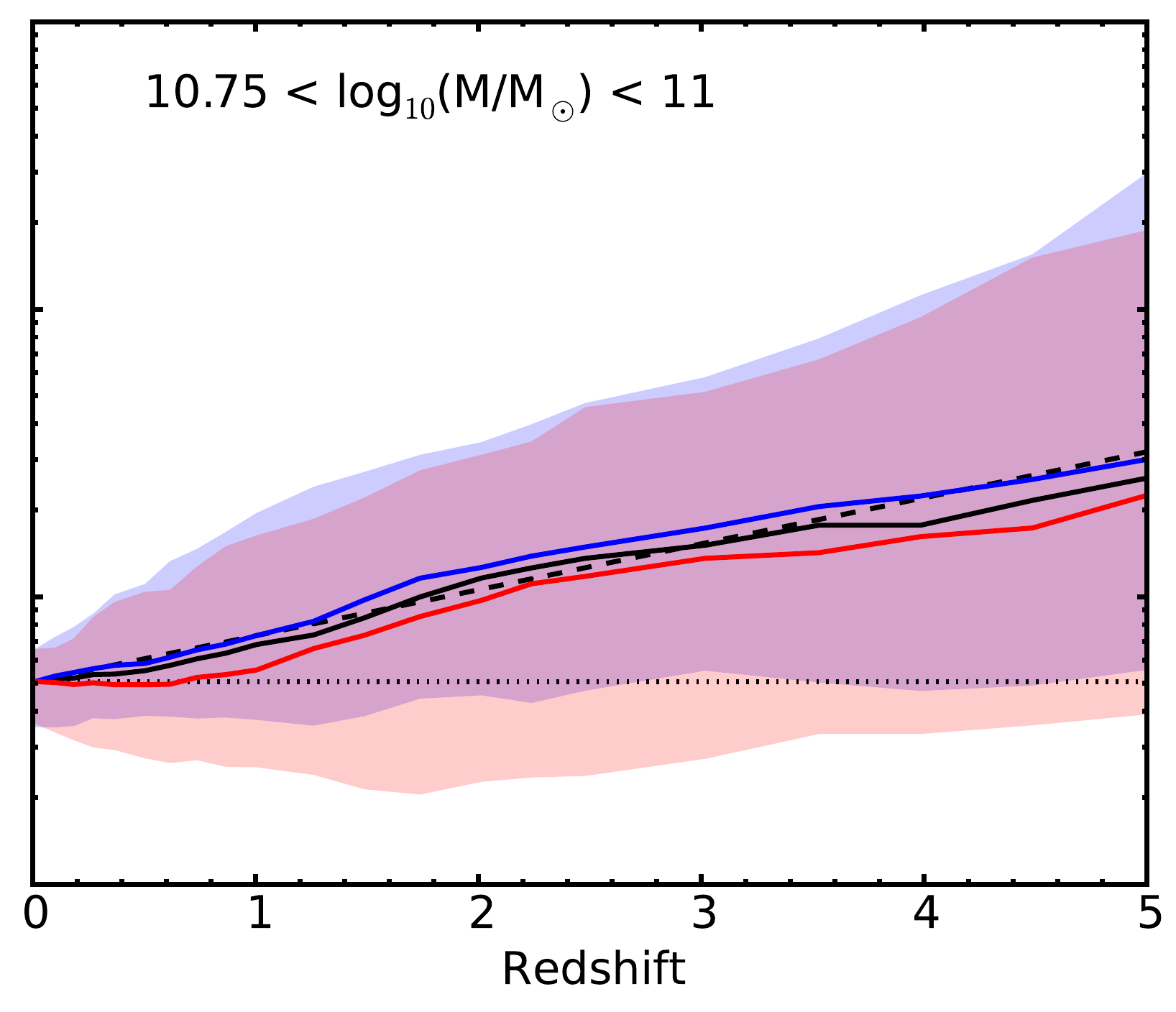}
\caption{The cumulative number density of main progenitors. All selections are the same as in Fig. \ref{Figure1}. The median main progenitor mass of the $z=0$ passive galaxies (solid red lines) tends to follow a constant cumulative number density at low redshift and an exponential increase in the cumulative number density parallel to the \protect{\citet{Behroozi13}} prescription at higher redshifts.}
\label{Figure2}
\end{figure*}

Fig. \ref{Figure1} shows the median mass history of main progenitors in three 0.25 dex wide mass bins at $z=0$. For all mass bins (also those not shown at lower and higher masses) a constant cumulative number density (dotted black curve) significantly overestimates the true median main progenitor mass (solid black curve) in the simulation. Typically there is already an 0.5 dex difference in mass at $z=2$. We confirm that this offset is adequately captured by the prescription of \citet{Behroozi13} of increasing the cumulative number density by $(0.16 \Delta z)$ dex (compare solid and dashed black curves). The fact that the EAGLE hydrodynamic simulation gives the same median main progenitor mass history as the abundance matching technique of \citet{Behroozi13}, shows that this is mainly a property of the dark matter halo merger tree, provided that the history of the GSMF is accurately captured by the simulation.

However, hydrodynamic processes do determine what kind of galaxy can be expected to be found in what kind of halo. The technique of finding representative main progenitors via cumulative number density matching assumes that galaxies of a certain mass share a common history. We know from observations that the redshift zero galaxy population is bimodal, with clear active (star forming) and passive galaxy populations \citep[e.g.][]{Strateva01}. We expect that the history of a typical passive galaxy differs from that of a typical active galaxy, remaining closer to its current stellar mass.

Indeed, in the EAGLE simulation the current active/passive status of a galaxy (defined using a sSFR cut at $\rm{10^{-11}/yr}$), is highly predictive for its median main progenitor mass history. The blue and red curves in Fig. \ref{Figure1} show the median main progenitor mass of the $z=0$ active, respectively passive, galaxy populations. The three mass bins cover the interesting region that goes from no difference between active/passive at $10.75<log_{10}(M/{\rm M_{\odot}})<11$ and higher, via a significant difference at Milky Way-like masses $10.5<log_{10}(M/{\rm M_{\odot}})<10.75$, towards a large difference at $10.25<log_{10}(M/{\rm M_{\odot}})<10.5$ and lower. These differences can be of the same order as those between a constant cumulative number density and the \citet{Behroozi13} prescription, roughly 0.5 dex at $z=2$. The $\rm{10^{th}-90^{th}}$ percentile blue and red shaded regions show that there is also a large variation in main progenitor masses. A recent study by \citet{Terrazas16} reports a comparable 0.35 dex difference at $z=2$ between the median main progenitor masses of active and passive $10.7<log_{10}(M/{\rm M_{\odot}})<10.9$ galaxies in the semi-analytic model of \citet{Henriques15}.

Fig. \ref{Figure2} shows the median cumulative number density for the same galaxy samples as in Fig. \ref{Figure1}. At a given redshift, the comoving cumulative density refers to the number density of galaxies with a stellar mass larger than or equal to the median main progenitor mass of the indicated galaxy sample, which is selected at $z=0$. We see that the median main progenitor of the active galaxies follows the exponential cumulative number density increase of \citet{Behroozi13}, but the median main progenitor of the passive galaxies first evolves along a track of constant cumulative number density up to $z \sim 1.5$, after which it follows the same exponential trend as the main progenitors of active galaxies, albeit at an offset which would correspond to a more massive active galaxy at redshift zero.

\begin{figure}
\includegraphics[width=1.0\columnwidth]{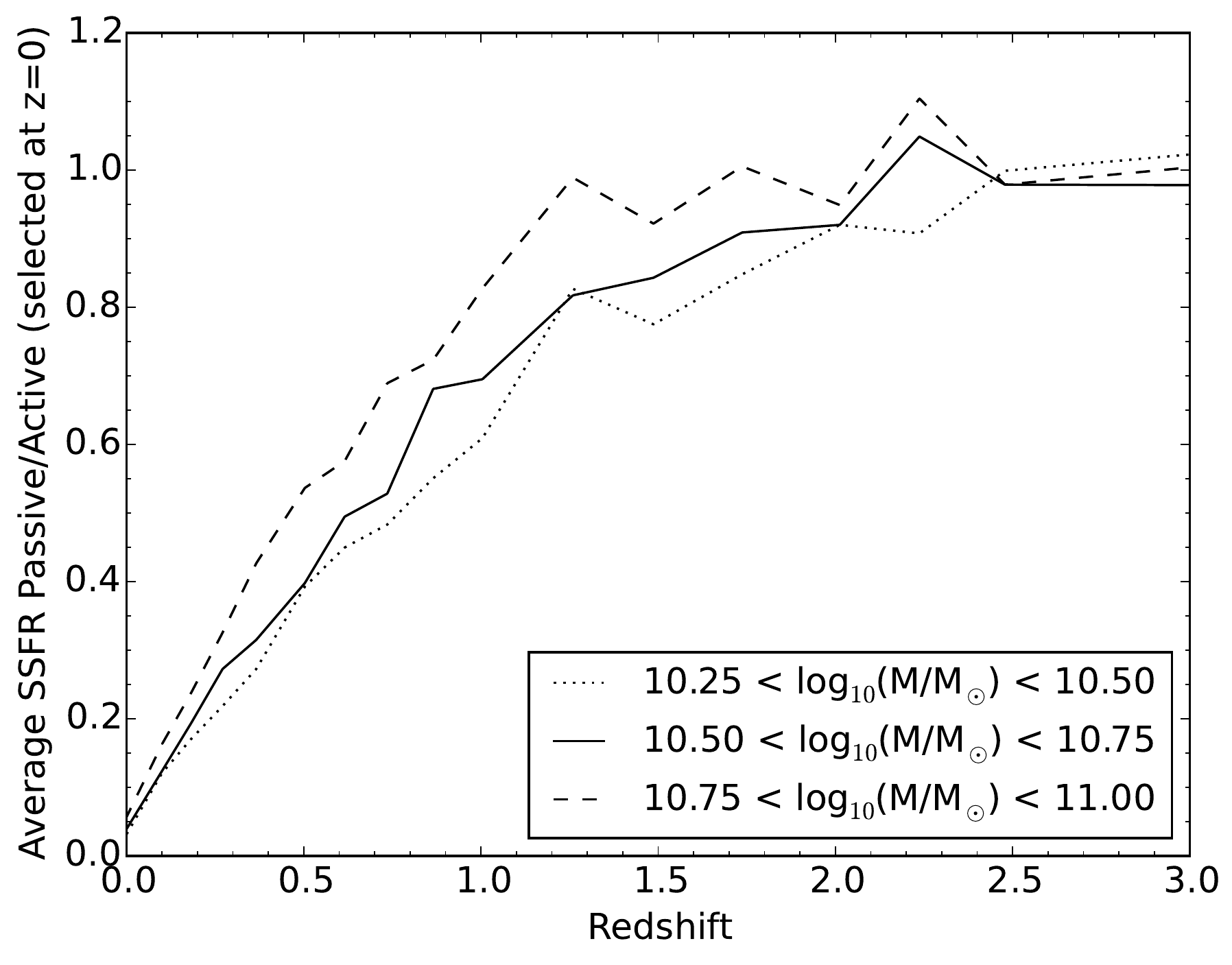}
\caption{The average specific star formation rate of the main progenitors of the passive $z=0$ galaxy sample divided by that of the active galaxy sample, as a function of redshift. The three curves are for three different, 0.25 dex wide, stellar mass bins.}
\label{Figure3}
\end{figure}

This behaviour can be explained by comparing the evolution of the sSFR of the main progenitors of the active versus passive $z=0$ galaxy selections. Fig. \ref{Figure3} shows the ratio of the average sSFRs of the passive and active galaxy selections in different  $z=0$ mass bins. We see that the main progenitors of passive galaxies have a reduced sSFR with respect to the main progenitors of active galaxies up to $z\approx1.25$ for the highest-mass bin and up to $z\approx2.5$ for the lowest-mass bin. Higher-mass passive galaxies have on average quenched later. Although not all the stellar mass growth can be attributed to the sSFR of the main progenitor, since dry mergers also contribute, the integrated effect of this sSFR difference between the active and passive samples plays a large part in driving the difference in median main progenitor mass and the corresponding difference in cumulative number density. The redshift range over which the passive main progenitors in Fig. \ref{Figure2} follow a constant cumulative number density roughly agrees with the redshift range in Fig. \ref{Figure3} for which the sSFR is reduced. In this same redshift range EAGLE matches the observed passive fraction as a function of galaxy stellar mass quite well (see Fig. 6, \citeauthor{Furlong15} \citeyear{Furlong15}).

\begin{figure}
\includegraphics[width=0.495\textwidth]{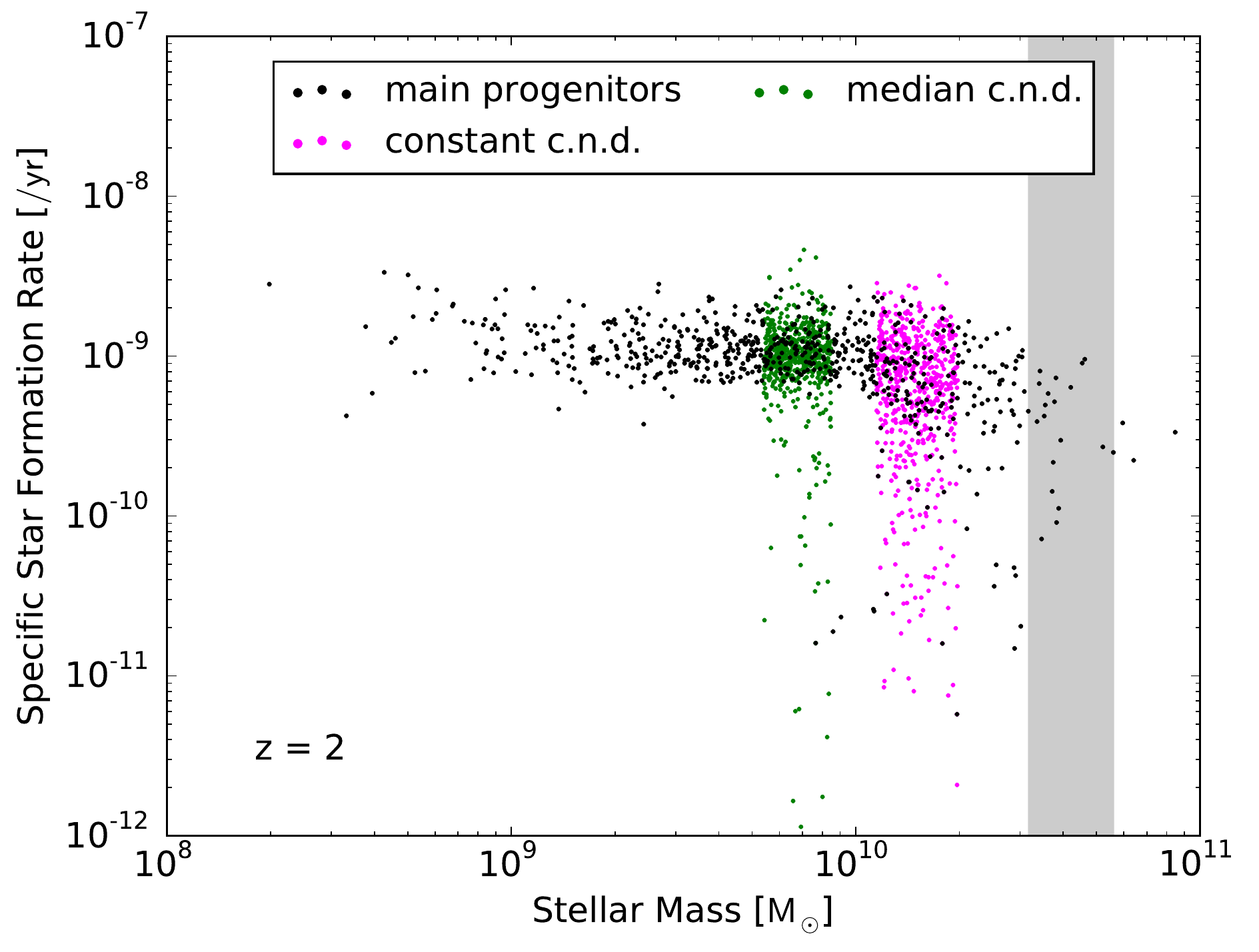}
\includegraphics[width=0.495\textwidth]{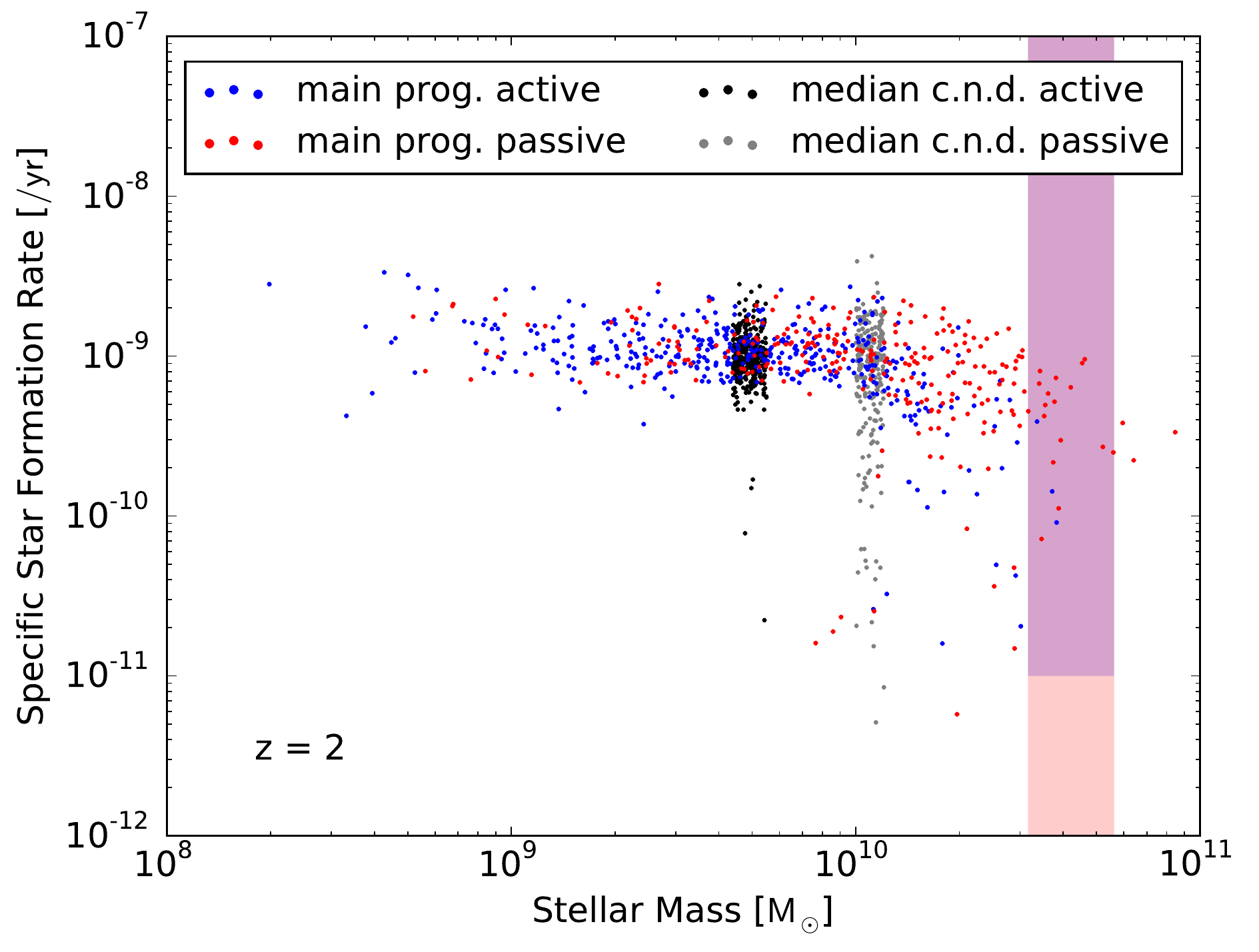}
\caption{The specific star formation rate versus stellar mass of the main progenitors at a redshift of 2 of the galaxy sample selected at redshift zero to have a stellar mass of $10.5<log_{10}(M/{\rm M_{\odot}})<10.75$ (indicated by the grey band). In the top panel, black dots denote the true main progenitors, magenta dots denote the galaxy sample that would be selected at $z=2$  based on a constant median cumulative number density and green dots denote the galaxy sample that would be selected using the true median main progenitor stellar mass, which gives similar results as the prescription of \citet{Behroozi13}. In the bottom panel a redshift zero selection is made on active/passive galaxies, denoted by the blue/red bands. Blue and red dots denote the true main progenitors of the active respectively passive galaxies. Black and grey dots denote the galaxy sample that would be selected at $z=2$ using the correct median main progenitor mass separately for the active and passive sample.}
\label{Figure4}
\end{figure}

In order to get a better insight into the difference between the true main progenitor galaxy sample and that obtained with different cumulative number density matching techniques, Fig. \ref{Figure4} shows a snapshot at redshift 2 of the two most relevant galaxy properties: the stellar mass and the sSFR, for the main progenitors of galaxies with Milky Way-like masses at redshift zero. The top panel shows the significant difference between the constant cumulative number density technique and that obtained by using the real median cumulative number density, which gives results that are very similar to the  \citet{Behroozi13} prescription. The bottom panel shows the difference when using the real median cumulative number density for active and passive galaxies separately. At $z=2$ the stellar mass range of the main progenitors, which was 0.25 dex at redshift zero, already spans several orders of magnitude. Since the EAGLE galaxies follow a relatively tight stellar mass versus dark matter mass relation, the same applies to the halo mass. Hence, any technique that does not sample a representative spread in stellar mass, will select a rather unrepresentative sample of main progenitors. Moreover, for such a wide distribution, the median and the mean stellar mass will differ substantially.

Depending on which galaxy property is studied, one might need a different cumulative number density matching technique. For example, a study like that of \citet{Dokkum13}, which uses a cumulative number density matching technique to observationally assess the radial stellar mass buildup of Milky Way-like galaxies, would benefit from using a cumulative number density based on the average stellar mass, or even better, including a representative variation in mass.

\begin{figure}
\includegraphics[width=1.0\columnwidth]{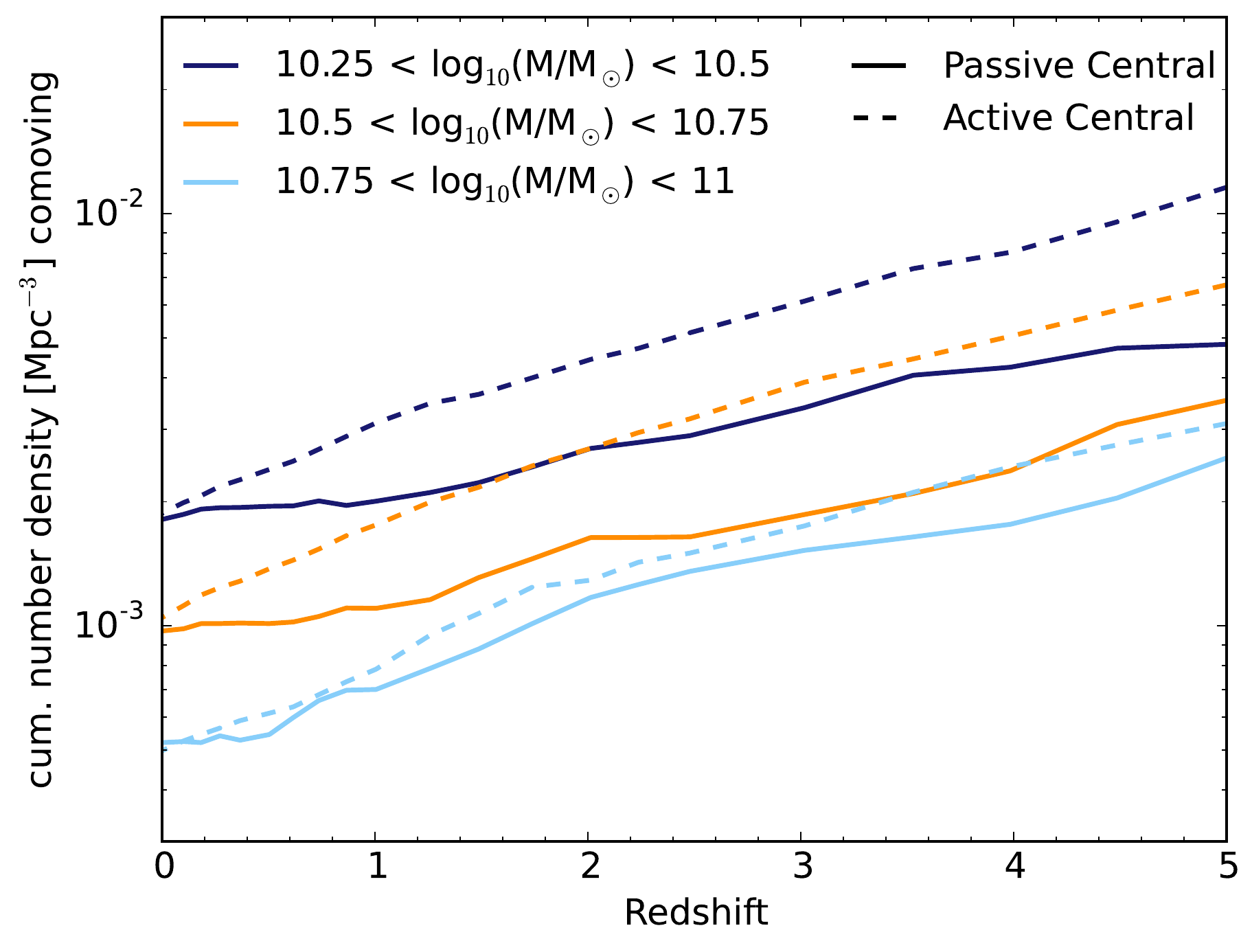}
\includegraphics[width=1.0\columnwidth]{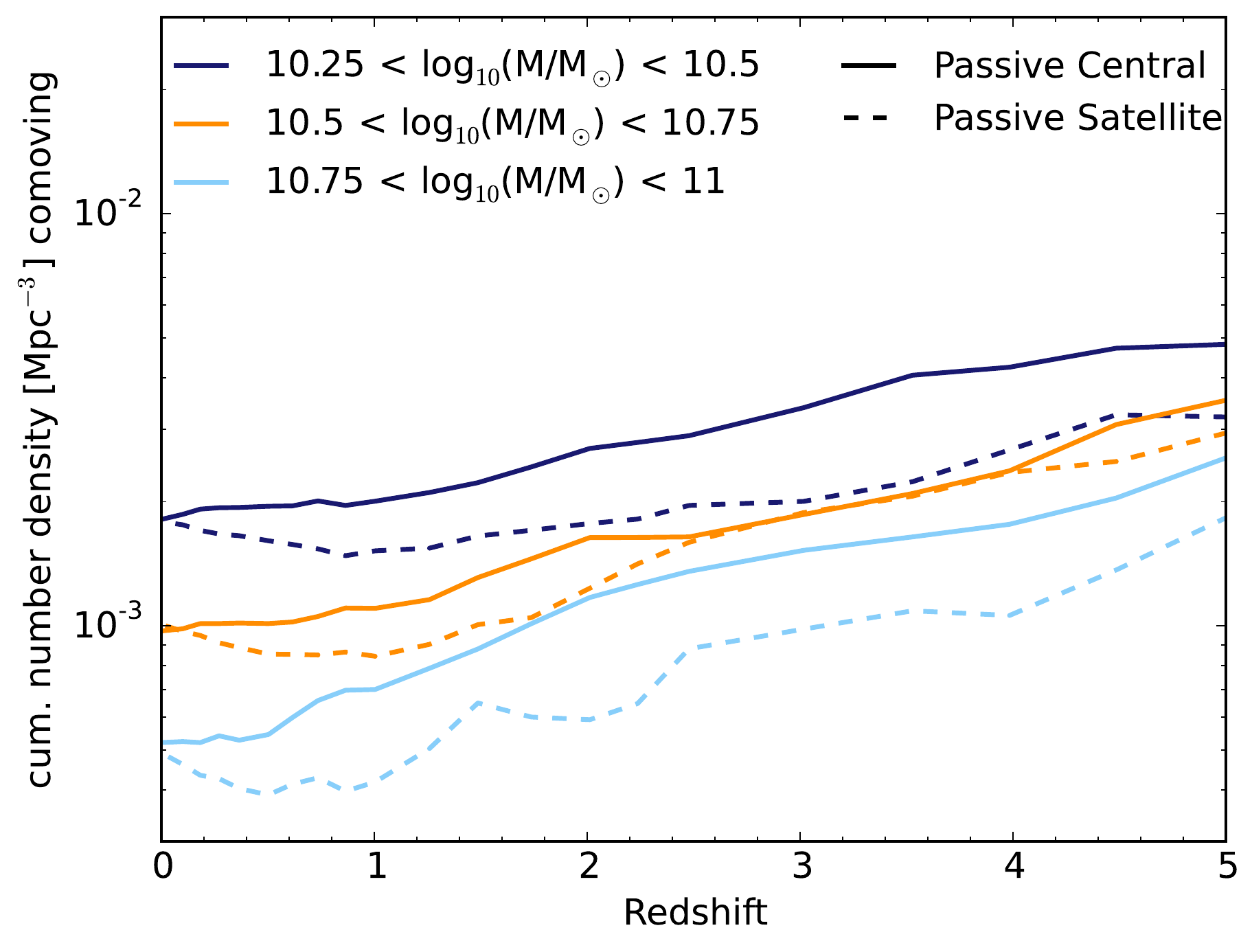}
\caption{The median cumulative number density of the main progenitors of $z=0$ galaxies in different mass bins (indicated by different colours), separately for passive and active centrals (top panel) and for passive centrals and satellites (bottom panel). The top panel shows a similar difference between active and passive cumulative number densities for central galaxies as Fig. \ref{Figure2} showed for all galaxies. The bottom panel shows that the main progenitors of passive satellites tend to be more massive than those of passive centrals of the same mass.}
\label{Figure5}
\end{figure}

Apart from discriminating between the progenitors of active and passive galaxies, an obvious other selection criterion is on satellite- versus central galaxies. Fig. \ref{Figure5} (top panel) shows that selecting only the main progenitors of central galaxies gives a similar dependence on the active/passive state as for all galaxies in Fig. \ref{Figure2}, so the difference in mass between the main progenitors of active and passive galaxies is not mainly caused by the quenching of satellite galaxies. Fig. \ref{Figure5} (bottom panel) shows however that the main progenitors of passive satellites tend to be more massive than those of passive centrals. This difference could be due to earlier quenching of satellite galaxies and/or the stripping of stars from satellite galaxies. There is no similar difference in mass between the main progenitors of active centrals and satellites (not shown).

Our results indicate that the progenitor masses are systematically offset for passive galaxies by an amount similar to the correction calculated by \citet{Behroozi13}. In addition, the properties of the progenitors are systematically correlated over time (e.g. SFR and hence central density, Sersic profile, etc.). This means that, in the absence of a procedure to correct for these correlations, the method has only limited applicability. At the highest masses $(n < 5\times10^{-4} \times10^{0.16 z} \, {\rm Mpc^{-3}})$ the method works best.

\section{Conclusions}
\label{SectionConclusions}

We use the EAGLE hydrodynamical simulation to study the accuracy of the cumulative number density matching technique in creating a representative sample of main progenitor galaxies as a function of redshift. The EAGLE simulation is well suited to study this question, because it reproduces the evolution of the galaxy stellar mass function and has a representative population of passive and active galaxies. Our main findings are as follows:

\begin{itemize}
\item Using a constant cumulative number density prescription to find typical main progenitors of redshift zero galaxies neglects mergers and significantly overestimates the median progenitor mass. The error is $\approx$ 0.5 dex at $z=2$ for galaxies in the stellar mass range $10^{10}{\rm M_{\odot}}<M<10^{11}{\rm M_{\odot}}$ and  $\approx0.25$ dex for $M>10^{11}\rm{M_{\odot}}$ (Fig. \ref{Figure1}).
\item The prescription of \citet{Behroozi13} to increase the cumulative number density by 0.16 dex per $\Delta z$ accurately captures the evolution of the median main progenitor stellar mass. As \citet{Behroozi13} used a dark mater only simulation, this shows that the evolution of the median main progenitor mass is mainly set by the properties of the halo merger tree (Fig. \ref{Figure2}).
\item The main progenitor masses of $z=0$ galaxies that are less massive than $\rm{10^{10.75}M_{\odot}}$ critically depend on the current star formation rate. At $z=2$ the median main progenitor mass of passive galaxies ($sSFR<10^{-11}/{\rm yr}$ at $z=0$) is $\approx 0.5$ dex higher than that of active galaxies (Figs. \ref{Figure1},\ref{Figure2}).
\item The difference between the median main progenitor mass of active and passive galaxies (or equivalently the difference in median cumulative number density) increases gradually up to $z\approx2$. Out to this same redshift we see a reduced average sSFR for the main progenitors of passive galaxies compared to those of active galaxies (Fig. \ref{Figure3}).
\item The mass difference between the main progenitors of passive and active galaxies persists if we only select central galaxies (Fig. \ref{Figure5}).
\item The large difference between main progenitor masses of passive and active galaxies calls for an inclusion of an sSFR distinction in the cumulative number density matching technique.
\item The spread in main progenitor masses already spans several orders of magnitude at $z=2$ (Fig. \ref{Figure4}). Therefore, it is imperative to include the variation in the main progenitor stellar masses (or cumulative number densities) in observational studies that use the cumulative number density matching technique. For such a wide distribution the average (e.g. main progenitor density profile) is expected to be different from the median.
\item The main progenitors of passive satellites tend to be more massive than those of passive centrals in the same $z=0$ mass range (Fig. \ref{Figure5}). A similar distinction between the main progenitor masses of active centrals and active satellites does not exist.
\end{itemize}

Hence, the cumulative number density matching technique should discriminate between the main progenitors of active and passive galaxies in order to obtain a reasonable accuracy. Exactly how to implement this is not clear. One could base such a method on a simulation like EAGLE and device a main progenitor selection scheme based on both stellar mass and sSFR. This would however negate
the initial appeal of the method: that it only depends on observations (and on the dark matter merger tree in the case of the \citeauthor{Behroozi13} \citeyear{Behroozi13} method).

\section*{Acknowledgements}

We  gratefully acknowledge support from the European Research Council under the European Union's Seventh Framework Programme (FP7/2007-2013) / ERC Grant agreement 278594-GasAroundGalaxies.

\bibliographystyle{mn2e} 
\bibliography{Bibliography}

\label{lastpage}
\end{document}